\documentclass[twocolumn,prl,amsmath,amssymb,floatfix,superscriptaddress,showpacs]{revtex4-1}
\usepackage{graphicx}% Include figure files
\usepackage{dcolumn}% Align table columns on decimal point
\usepackage{bm}% bold math
\usepackage{xspace}
\usepackage{hyperref}
\usepackage{color}

\def\rucl{$\alpha$-RuCl$_3$\xspace}
\begin{document}
% \draft command makes pacs numbers print

%\setlength{\topmargin}{0in} %%%set margin to be zero to have proper ps printout
%%%when uploaded to arxiv and aps, this has to be commented.
\title{Spin-wave excitations evidencing the Kitaev interaction in single crystalline $\alpha$-RuCl$_3$}
\author{Kejing~Ran}
\author{Jinghui~Wang}
\author{Wei~Wang}
\author{Zhao-Yang~Dong}
\affiliation{National Laboratory of Solid State Microstructures and Department of Physics, Nanjing University, Nanjing 210093, China}
\author{Xiao~Ren}
\affiliation{International Center for Quantum Materials, School of Physics, Peking University, Beijing 100871, China}
\author{Song~Bao}
\author{Shichao~Li}
\author{Zhen~Ma}
\author{Yuan~Gan}
\author{Youtian~Zhang}
\affiliation{National Laboratory of Solid State Microstructures and Department of Physics, Nanjing University, Nanjing 210093, China}
\author{J.~T.~Park}
\affiliation{Heinz Maier-Leibnitz Zentrum (MLZ), TU M\"{u}nchen, Garching D-85747, Germany}
\author{Guochu~Deng}
\author{S.~Danilkin}
\affiliation{ Australian Nuclear Science and Technology Organisation (ANSTO), New Illawarra Road, Lucas Heights, NSW 2234, Australia}
\author{Shun-Li~Yu}
\author{Jian-Xin~Li}
\altaffiliation{jxli@nju.edu.cn}
\author{Jinsheng~Wen}
\altaffiliation{jwen@nju.edu.cn}
\affiliation{National Laboratory of Solid State Microstructures and Department of Physics, Nanjing University, Nanjing 210093, China}
\affiliation{Collaborative Innovation Center of Advanced Microstructures, Nanjing University, Nanjing 210093, China}

\date{\today}

\begin{abstract}
Kitaev interactions underlying a quantum spin liquid have been long sought, but experimental data from which their strengths can be determined directly is still lacking. Here, by carrying out inelastic neutron scattering measurements on high-quality single crystals of $\alpha$-RuCl$_3$, we observe spin-wave spectra with a gap of $\sim$2~meV around the M point of the two-dimensional Brillouin zone. We derive an effective-spin model in the strong-coupling limit based on energy bands obtained from first-principle calculations, and find that the anisotropic Kitaev interaction $K$ term and the isotropic antiferromagentic off-diagonal exchange interaction $\Gamma$ term are significantly larger than the Heisenberg exchange coupling $J$ term. Our experimental data can be well fit using an effective-spin model with $K=-6.8$~meV and $\Gamma=9.5$~meV. These results demonstrate explicitly that Kitaev physics is realized in real materials.

\end{abstract}


\maketitle
Quantum spin liquids (QSLs) are an exotic topological state of matter in which strong quantum fluctuations prevent conventional magnetic order from establishing down to zero temperature~\cite{Anderson1973153}. Examples of such states have been proposed in geometrically frustrated quantum magnets having a small spin of $S=1/2$, where the isotropic Heisenberg interaction $J$ cannot be satisfied simultaneously among different sites~\cite{nature464_199,nature492_406}. In these materials, the degeneracy is so large that the exact state of the system has been challenging to determine. The Kitaev QSL provides an alternative~\cite{aop321_2}; it is realized in the exactly solvable Kitaev spin model, which has a bond-dependent anisotropic exchange interaction $K$ with an intrinsic frustration of the spin on a single site~\cite{PhysRevLett.98.247201,aop321_2}. The Kitaev QSL can host topological order and non-Abelian statistics~\cite{aop321_2,prl105_027204,PhysRevLett.108.127203,PhysRevLett.112.207203,PhysRevB.92.115127},  the latter property being associated with proposals for fault-tolerant quantum computation~\cite{Kitaev20032,RevModPhys.80.1083,aop321_2}. Therefore, investigating Kitaev spin liquids is of both fundamental and practical importance.

\begin{figure}[htb]
\centering
\includegraphics[width=0.98\linewidth,trim=0mm 10mm 5mm 35mm,clip]{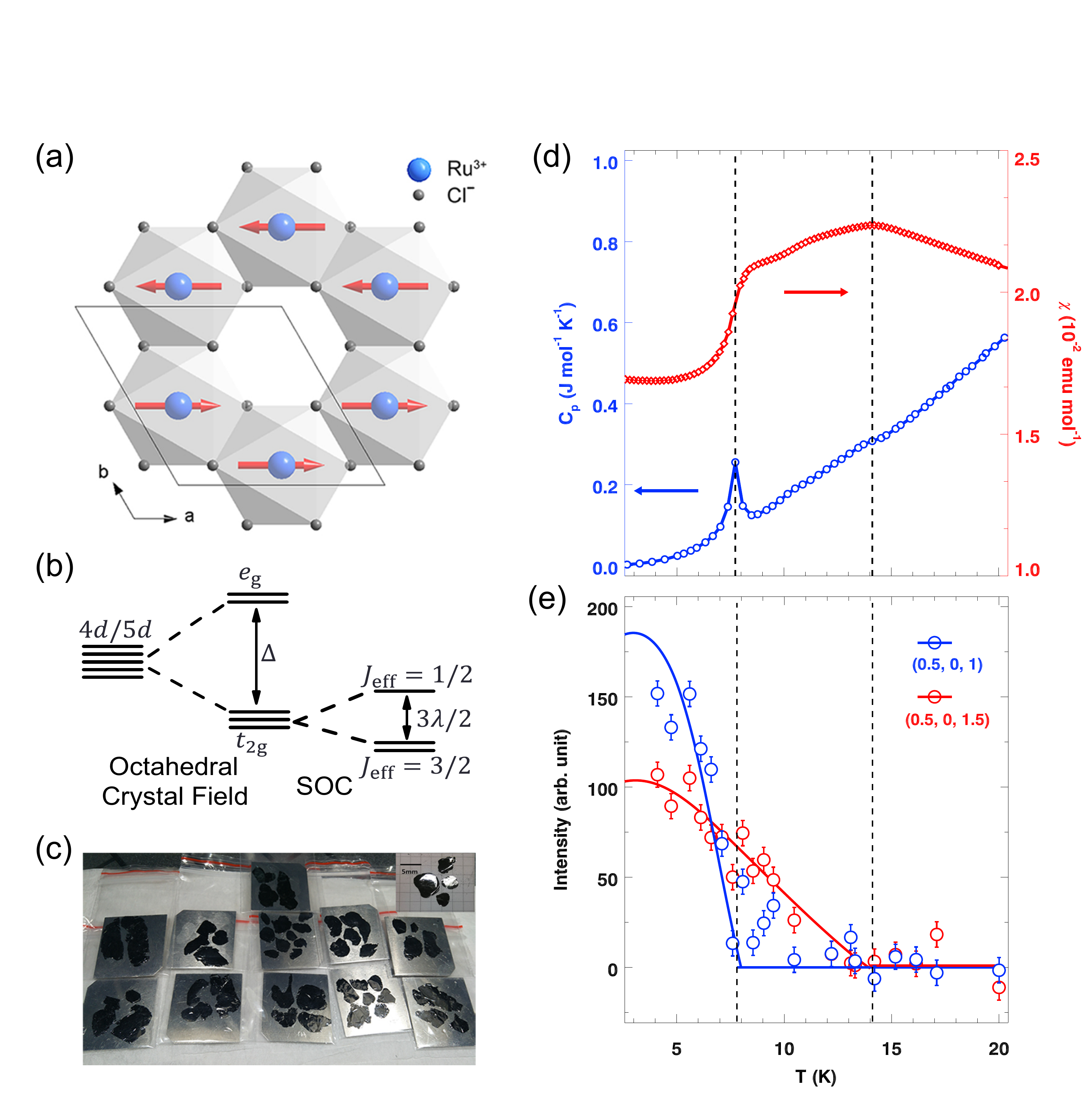}
\caption{\label{fig:structure}{(Color online) (a) One honeycomb layer of \rucl. Thick arrows represent one configuration of the three 120$^\circ$-twinned magnetic domains for the zigzag order. (b) Effective spin $J_{\rm{eff}}=1/2$ configuration of the 4$d$ or 5$d$ orbitals subject to an octahedral crystal field $\Delta$, an SOC $\lambda$, and electron correlations. (c) Single crystals with a typical size of $5\times5\times0.5$~mm$^3$ for each piece, glued on (both faces of) the aluminum plates. (d) Heat capacity measured in zero magnetic field (left axis) and susceptibility measured in a 0.5-T field applied parallel to the $a$-$b$ plane (right axis). (e) Temperature dependence of the integrated intensities of the magnetic peaks (0.5,\,0,\,1) and (0.5,\,0,\,1.5), obtained by fitting $L$ scans through the peaks. Dashed lines indicate magnetic transition temperatures. Lines through data are guides to the eye. Throughout the paper, errors represent one standard deviation.}}
\end{figure}

Engineering of Kitaev interactions in real materials was first proposed in iridates~\cite{prl102_017205}. In these materials, iridium and oxygen ions form edge-sharing octahedra~\cite{RevModPhys.82.53}, same as that sketched in Fig.~\ref{fig:structure}(a), but with Ru$^{3+}$ and Cl$^{-}$ replaced by Ir$^{4+}$ and O$^{2-}$ respectively. As illustrated in Fig.~\ref{fig:structure}(b)~\cite{PhysRevLett.110.076402,Kim1329,PhysRevLett.114.096403,doi:10.1146/annurev-conmatphys-031115-011319}, a strong octahedral crystal field splits the five $d$ orbitals into triply-degenerate $t_{2g}$ and doubly-degenerate $e_g$ states. Due to the substantial spin-orbital coupling (SOC) of 5$d$ electrons, the degeneracy of the $t_{2g}$ states is lifted by the formation of $J_{\rm{eff}}=3/2$ and 1/2 bands, with a total angular momentum of 3/2 and 1/2 respectively. Since the $J_{\rm{eff}}=1/2$ band is narrow, a Mott gap opens in the presence of modest electron correlations, and the system is driven to a Mott insulating state with an effective spin of $1/2$. In these systems, the magnetic interaction of the effective spin is intrinsically anisotropic and frustrated due to the spatially anisotropic nature of the $d$ orbitals and the SOC. The delicate bond configuration of the octahedra makes this interaction prominent, naturally fitting Kitaev's proposal~\cite{prl102_017205,prl105_027204,np11_462,prl110_097204}.
More recently, the layered honeycomb-lattice compound \rucl [see Fig.~\ref{fig:structure}(a)] has been suggested to be another material where Kitaev physics may be applicable~\cite{PhysRevB.91.241110,PhysRevLett.114.147201,PhysRevB.90.041112}. For \rucl with 4$d$ electrons, the SOC constant $\lambda$ of $\sim$0.13~eV~\cite{PhysRevB.93.075144} is weaker than that of the iridates with $\lambda\approx0.4$~eV~\cite{doi:10.1146/annurev-conmatphys-031115-011319}, but since the band is also narrower due to stronger electron correlations, the same $J_{\textnormal{eff}}=1/2$ state as illustrated in Fig.~\ref{fig:structure}(b) can also be achieved~\cite{PhysRevB.53.12769,PhysRevLett.117.126403}. Indeed, \rucl has closer-to-ideal bond configurations so that Kitaev interactions dominate over the isotropic Heisenberg exchange coupling~\cite{J19670001038,PhysRevB.94.085109,PhysRevB.94.064435}.
The most powerful approach to identify magnetic interactions is to perform inelastic neutron scattering (INS) measurements on single crystals to explore their dynamical spin response. However, iridium is a strong absorber of neutrons, making neutron scattering experiments challenging~\cite{PhysRevLett.108.127204}; as for \rucl, there has been only one report of INS measurements on powder samples~\cite{nm15_733}.

In this Letter, by performing INS measurements on high-quality single crystals of \rucl, we map out the low-energy spin-wave excitation spectra. By using results obtained from first-principle calculations and fitting the experimental data with a minimal effective-spin model, we determine the two leading interactions, including a ferromagnetic Kitaev $K$ term of $-6.8$~meV, and an antiferromagnetic symmetric off-diagonal $\Gamma$ term of $9.5$~meV. Our results clearly indicate that the anisotropic Kitaev coupling plays a key role in the exotic physics of \rucl. 

Single crystals of \rucl were grown by the chemical-vapor-transport method. The crystals are plate like, as shown in Fig.~\ref{fig:structure}(c), with a typical mass of 20~mg for each piece. Specific heat and susceptibility measurements were conducted in a Physical Property Measurement System (PPMS) from Quantum Design. Neutron scattering measurements were performed on two triple-axis spectrometers, TAIPAN at ANSTO, and PUMA at MLZ~\cite{puma}, both using a fixed-final-energy mode with $E_f=14.7$~meV. Measurements on both instruments were performed under double-focusing conditions for both the monochromator and analyzer. To maximize neutron flux, no collimator was used in the measurements. On TAIPAN, we put a pyrolitic graphite (PG) filter after the sample to reduce higher-order neutrons; on PUMA, we put two PG filters after the sample for the same purpose, and one sapphire filter before the monochromator to suppress epithermal neutrons. Under these conditions, the energy resolution (half width at half maximum) for both instruments was $\sim$1~meV at the Bragg position. For the measurements on TAIPAN, we co-aligned 50 single crystals weighing about 1~g in total using a Laue X-ray diffractometer. Crystals were glued to both faces of the aluminum plates using hydrogen-free cytop grease. Some of the plates before assembling are shown in Fig.~\ref{fig:structure}(b). These crystals were well aligned such that the overall mosaic was $\sim$1.4$^\circ$ as determined from the rocking scan through the (300) peak. The samples on the plates were oriented such that the scattering plane was ($H0L$), where we obtained the neutron diffraction data. As for the measurements on PUMA, we used 80 single-crystal pieces weighing $\sim$1.5~g with an overall mosaic of $\sim$4.7$^\circ$ measured at (300). The measurements on PUMA were conducted in the $(HK0)$ plane, where we obtained the INS data. The data were analyzed and plotted with the Interactive Data Language programming environment. There have been some discrepancies on whether the low-temperature structure belongs to the monoclinic C2/m~\cite{PhysRevB.92.235119,PhysRevB.93.134423}, or trigonal P3$_1$12 group~\cite{sfi12_152,PhysRevB.93.155143}. Recently, it has been reported that a phase transition from monoclinic to trigonal structure occurs at $150$~K upon cooling~\cite{PhysRevB.91.094422}. Throughout the paper, we use P3$_1$12 notation with $a=5.96$~\AA\ and $c=17.2$~\AA. Such a structure has been commonly adopted for neutron measurements~\cite{PhysRevB.91.144420,PhysRevB.90.041112,nm15_733}. The wave vector $\bm{Q}$ is expressed as ($HKL$) reciprocal lattice unit (r.l.u.) of $(a^{*},\,b^{*},\,c^{*})=(4\pi/\sqrt{3}a,\,4\pi/\sqrt{3}b,\,2\pi/c)$. 

We have carefully characterized the crystals by energy-dispersive X-ray spectroscopy (confirming the composition), a Laue X-ray diffractometer (confirming the single crystallinity), and a PPMS. Heat capacity and susceptibility measured in the PPMS are presented in Fig.~\ref{fig:structure}(d), which show that there are two transitions in the samples. The transition temperatures of $\sim$8~K ($T_{\rm{N1}}$) and 14~K ($T_{\rm{N2}}$) almost coincide with the onsets of the intensities for the two magnetic Bragg peaks (0.5,\,0,\,1) and (0.5,\,0,\,1.5), respectively [shown in Fig.~\ref{fig:structure}(e)]. The two transitions have been reported in a number of works~\cite{PhysRevB.91.144420,PhysRevB.90.041112,PhysRevB.91.094422}, and have been interpreted as a result of ABC- and AB-type stacking arrangements of the honeycomb layers~\cite{nm15_733}. For samples with the ABC-type stacking structure alone, only the low-temperature transition has been observed~\cite{PhysRevB.93.134423}. The observed magnetic order with an in-plane wave vector of (0.5,\,0) instead of (0.5,\,0.5) is consistent with the zigzag order within the honeycomb layer, as shown in Fig.~\ref{fig:structure}(a). 

\begin{figure}[htb]
\centering
\includegraphics[width=0.98\linewidth,trim=42mm 57mm 20mm 82mm,clip]{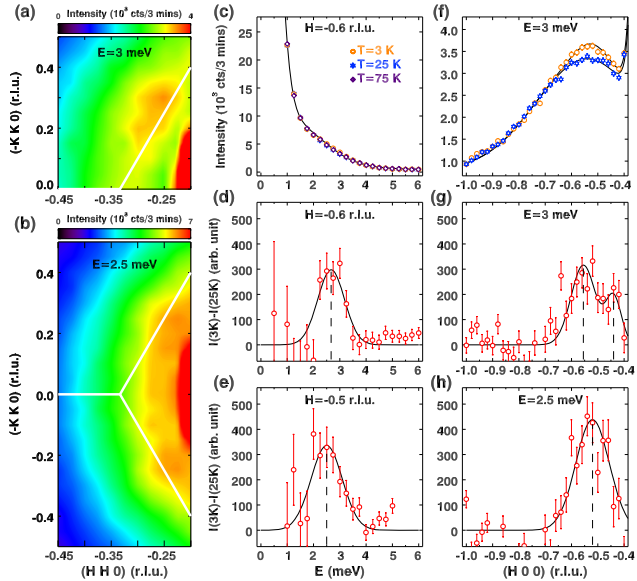}
{\caption{\label{fig:scans}{(Color online) (a) and (b) Contour maps at 3 and 2.5~meV respectively, measured at 3~K. Lines indicate the Brillouin zone boundary. (c) Energy scans at ($-0.6,\,0,\,0$) at various temperatures. (d) Difference between the 3- and 25-K scans of (c). (e) Same as (d) but at (-0.5,\,0,\,0). (f) Constant-energy scans with $E=3$~meV at 3 and 25~K. (g) Difference of the scans in (f). (h) Same as (g) but at $E=2.5$~meV. Lines through data are guides to the eye. Dashed lines indicate peak positions.}}}
\end{figure}

\begin{figure*}[htb]
\centering
\includegraphics[height=0.95\linewidth,angle=90,trim=25mm 0mm 70mm 10mm,clip]{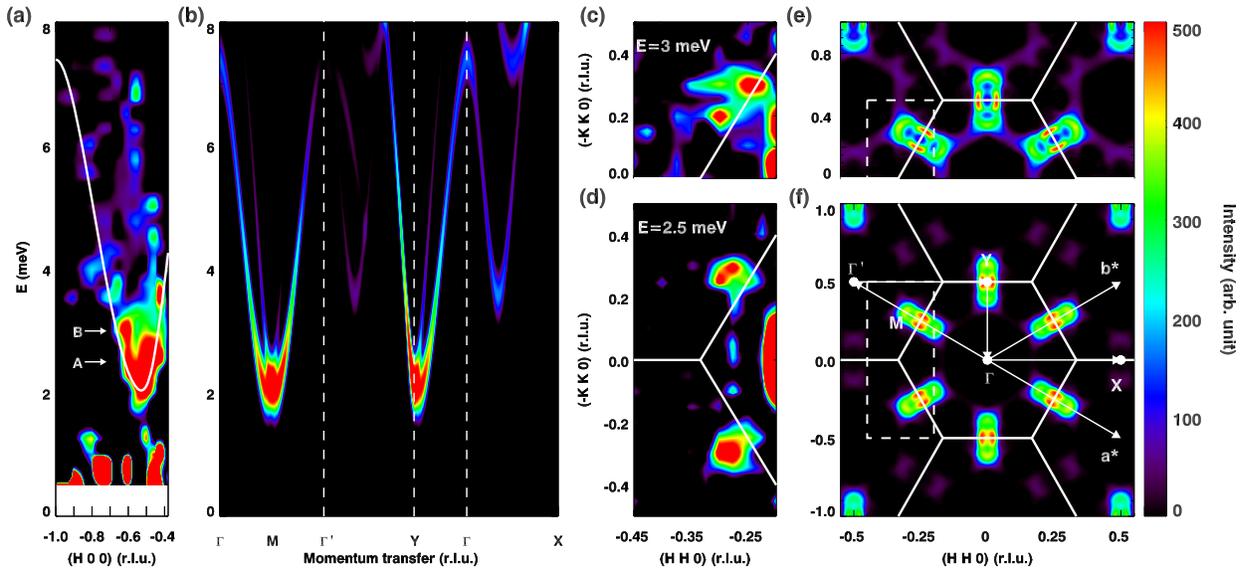}%%%lbrt
\caption{\label{fig:dispersion}{(Color online) (a) Magnetic dispersion along $[H00]$ obtained by plotting the differences of the energy scans with a step size of 0.25~meV at 11 $\bm{Q}$ points along the $\Gamma'$-$\Gamma$ path with $H$ ranging from -0.38 to -1~r.l.u.. From $H=-0.4$ to $-0.7$~r.l.u., the $H$ interval is 0.05~r.l.u., and from -0.7 to -1~r.l.u., the interval is 0.1~r.l.u.. The solid line is the calculated dispersion as shown in (b). Intensities below 1.5~meV are contaminated by nuclear scattering; intensities above 4~meV falling out of the dispersion are contaminated by low-angle scattering. The two arrows denote constant-energy cuts at 3 and 2.5~meV as plotted in (c) and (d). (b) Calculated spin-wave spectra along high-symmetry paths of the two-dimensional Brillouin zone as indicated in (f). Dashed lines denote the boundaries. (c) and (d) Differences of the 3- and 25-K energy cuts at 3 and 2.5~meV, respectively. (e) and (f) Calculated constant-energy cuts at $E=3$ and 2.5~meV, respectively. Dashed rectangles denote the areas covered by experimental data. In (c)-(f), we plot the results with two orthogonal axes [$HH0$] and [$-KK$0], and the Brillouin zone boundary is indicated by the solid lines. (c) [(d)] consists of 11 (21) scans with $(-KK0)$ ranging from 0 (-0.5) to 0.5 (0.5)~r.l.u., along the $[HH0]$ direction with a step size of 0.025~r.l.u..}}
\end{figure*}

In \rucl, magnetic layers are only weakly bonded with the van der Waals force along the $c$ axis, resulting in a quasi-two-dimensional nature of the magnetic interactions~\cite{PhysRevB.91.144420,PhysRevB.90.041112,nm15_733}. We therefore investigate the spin dynamics within the $a$-$b$ plane. There is a broad continuum as shown in Fig.~\ref{fig:scans}(a), (b), (c) and (f), on top of which features around the M point at temperatures below $T_{\rm{N1}}$ are evident, as shown in Fig.~\ref{fig:scans}(a) and (b). By subtracting the 25-K data (above $T_{\rm{N2}}$) from the 3-K data (below $T_{\rm{N1}}$), the differences become obvious, as shown in Fig.~\ref{fig:scans}(d), (e), (g) and (h). The resulting peaks, with positions indicated by the dashed lines, are dispersive. By plotting the differences of the energy scans at various $\bm{Q}$ values on the $\Gamma'$-$\Gamma$ path, we have obtained the dispersion along the [100] direction as shown in Fig.~\ref{fig:dispersion}(a); by plotting the differences of various $\bm{Q}$ scans at $E=3$ and 2.5~meV, we have obtained constant-energy cuts as shown in Fig.~\ref{fig:dispersion}(c) and (d).

In the dispersion shown in Fig.~\ref{fig:dispersion}(a), there is a gap of $\sim$2~meV at the M point, similar to INS results on powder samples~\cite{nm15_733}; however, it shows maximum intensity near the M point with an energy of $\sim$2.5~meV, and concave curvature around this point, both of which are different from the powder results~\cite{nm15_733}. The energy cuts in Fig.~\ref{fig:dispersion}(c) and (d) show that the intensities are concentrated around the M point, with that of the 2.5-meV cut in Fig.~\ref{fig:dispersion}(d) centering at ($\pm0.55$,\,0,\,0). We suspect that this off-M-point behavior is an artifact resulting from the data subtraction, as the scattering is more pronounced at low-$\bm{Q}$ positions before the subtraction. From the dispersion shown in Fig.~\ref{fig:dispersion}(a), the center of $H$ is determined to be about -0.53~r.l.u., which is the same as that obtained by fitting the $H$ scan shown in Fig.~\ref{fig:scans}(h). Comparing Fig.~\ref{fig:dispersion}(c) and (d), it is obvious that the 3-meV data are broader along both directions (parallel and perpendicular to $[H00]$) than the 2.5-meV data. A width of 0.20 and 0.14~r.l.u.\ along the $[H00]$ direction is determined from Fig.~\ref{fig:dispersion}(c) and (d), respectively. These values are close to the results of 0.22 and 0.16~r.l.u.\ for $E=3$ and 2.5~meV respectively, extracted from the dispersion shown in Fig.~\ref{fig:dispersion}(a). As discussed above, this material has a zigzag magnetic order with an in-plane wave vector of (0.5,\,0), $i.e.$, the M point. All of the behavior of this dispersion point to the conclusion that the magnetic excitations are spin waves arising from the zigzag magnetic order phase.

\begin{figure}[htb]
  \centering
  \includegraphics[width=0.98\linewidth,trim=35mm 55mm 30mm 165mm,clip]{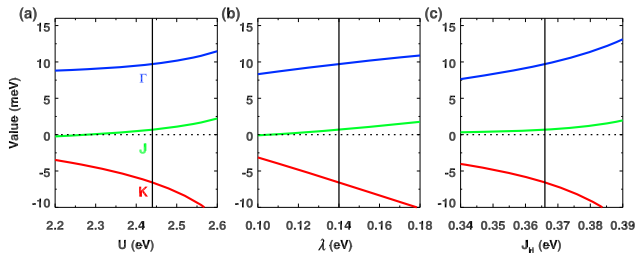}\\
  {\caption{\label{fig:fpcal}{(Color online) Calculated dependence~\cite{WWang} of $\Gamma$, $J$, and $K$ on Hubbard $U$ (a), SOC $\lambda$ (b), and Hund's coupling $J_{\rm H}$ (c). Vertical lines indicate one set of parameters with $U=2.44$~eV, $\lambda=0.14$~eV, and $J_{\rm H}$=0.37~eV.}}}
\end{figure}

The electronic structure of \rucl can be described by the five-orbital Hubbard model with an intermediate spin-orbital interaction~\cite{PhysRevB.93.214431,PhysRevB.91.241110}. For the Ru$^{3+}$ $4d^{5}$ electrons, the combined effects of the octahedral crystal field, SOC, and electron correlations lead to an effective picture of a hole residing on the $J_{\rm eff}=1/2$ Kramers doublet, as illustrated in Fig.~\ref{fig:structure}(b)~\cite{PhysRevB.53.12769,PhysRevLett.117.126403,PhysRevB.91.241110}.
In the large Hubbard $U$ limit, the effective Hamiltonian for the pseudospin is obtained by projecting the corresponding spin-orbital model onto the Kramers doublet~\cite{prl110_097204,prl102_017205}. In this context, we determine the exchange interactions in the effective-spin model by using tight-binding parameters obtained from first-principle calculations~\cite{WWang}, and the results are summarized in Fig.~\ref{fig:fpcal}. In the parameter ranges relevant to this system~\cite{PhysRevB.90.041112,PhysRevB.93.075144,PhysRevB.53.12769,PhysRevB.91.241110,PhysRevB.93.214431}, we find that the Kitaev interaction $K$ term is ferromagnetic, and the off-diagonal exchange $\Gamma$ term is antiferromagnetic, both of which are significantly larger than the Heisenberg $J$ term, as shown in Fig.~\ref{fig:fpcal}. The small $J$ value is likely to be a result of the virtual intraband hopping between the $J_{\rm eff}=1/2$ states, and the interband hopping between the $J_{\rm eff}=1/2$ and $3/2$ states, both of which contribute to the $J$ term~\cite{PhysRevB.93.214431}, nearly cancelling out~\cite{WWang}. We thus consider the following $K$-$\Gamma$ effective-spin model:
\begin{equation*}\label{Hamiltonian}
  H=\sum_{\langle ij\rangle \in\alpha\beta(\gamma)}[KS^\gamma_i S^\gamma_j+\Gamma(S^\alpha_i S_j^\beta+S^\beta_i S^\alpha_j)],
\end{equation*}
where $\alpha\beta(\gamma)$ labels one bond with $\alpha$, $\beta$ and $\gamma$ being the spin directions.

To obtain parameters which fit the INS data, we calculate the dynamical spin-spin correlation function utilizing linear spin-wave theory~\cite{PhysRevB.68.134424,Muniz01082014}. In the calculations, we have summed the intensities over the three 120$^\circ$-twinned magnetic domains, taking into account polarization factors and the instrumental resolution. Numerical results for dispersions along high-symmetry paths are presented in Fig.~\ref{fig:dispersion}(b). To compare with experimental data, we plot the calculated results along the $\Gamma'$-$\Gamma$ path as a solid line in Fig.~\ref{fig:dispersion}(a). With a ferromagnetic Kitaev interaction $K=-6.8$~meV, and an antiferromagnetic $\Gamma=9.5$~meV, the calculated dispersion agrees well with the data. Due to weak intensities, we have observed only one branch of the spin waves. The calculation shows an intensity maximum near the M point, in line with the experimental results. The spin-wave velocity can be increased slightly to match the data with $K=-7.2$~meV and $\Gamma=9.5$~meV. However, this combination reduces the gap size to 1.5~meV. One can increase $\Gamma$ as well to keep $|K/\Gamma|$ constant, and thus maintain the gap of 2~meV. We find that $|K/\Gamma|$ in the range of 0.65 to 0.8 gives gap sizes of about 2~meV. However, a noticeable change in $\Gamma$ requires significant changes of both $U$ and $\lambda$, as indicated in Fig.~\ref{fig:fpcal}(a) and (b). Overall, the choice of the above parameter set is reasonable since the resulting fit well captures the maximum intensities, gap value, and intensities above 4~meV. Energy cuts at 3 and 2.5~meV for the calculated dispersions are shown in Fig.~\ref{fig:dispersion}(e) and (f) respectively, which agree with the experimental data shown in Fig.~\ref{fig:dispersion}(c) and (d) qualitatively. 

To describe the anisotropic magnetic interactions in iridates and \rucl, a Heisenberg-Kitaev ($J$-$K$) model is commonly adopted, where $K$ plays a dominant role~\cite{prl105_027204,prl110_097204,PhysRevB.94.064435}. However, it has been found that this model has some difficulties in reproducing the zigzag magnetic order~\cite{PhysRevB.88.035107}, and a couple of alternatives have been proposed, including modified $J$-$K$ models with long-range terms~\cite{PhysRevB.84.180407,PhysRevLett.108.127204}, or even a completely different quasimolecular-orbital model~\cite{PhysRevLett.109.197201}. Models with additional terms such as the $\Gamma$ term have also been put forward~\cite{PhysRevB.93.214431,PhysRevLett.112.077204,PhysRevB.90.155126}. 
As discussed above, to mimic the gap of 2~meV, a finite $\Gamma$ (comparable to $K$) is necessary (see also Ref.~\cite{nm15_733}). Furthermore, in Ref.~\cite{PhysRevLett.112.077204}, it has been shown theoretically that for the zigzag ordered state, a major portion of the parameter space is filled by a large ferromagnetic $K$ and antiferromagnetic $\Gamma$ term, but nearly zero $J$ term, supporting our conclusions.

In summary, we have presented neutron scattering results on single crystals of \rucl showing that the excited states associated with the zigzag magnetic order can be well described by a minimal $K$-$\Gamma$ effective-spin model. These results make a strong case that \rucl is a prime candidate in realizing Kitaev physics. As for a Kitaev QSL, the only two known material systems by now, iridates and \rucl, indeed have some obvious drawbacks---they both show magnetic order, albeit weakly~\cite{PhysRevB.90.041112,PhysRevB.85.180403,PhysRevB.87.144405,PhysRevB.87.140406}.
Recently, it has been shown that in \rucl the magnetic order can be fully suppressed by an external magnetic field applied parallel to the plane~\cite{PhysRevB.91.180401,PhysRevB.92.235119,PhysRevB.91.094422}. It should be intriguing to study the material's spin dynamics under such a field. 

KJR, JHW, and WW contributed equally to the work. Work at Nanjing University was supported by NSFC Nos. 11374143, 11674157, 11190023, 11374138 and 11204125, and National Key Projects for Research \& Development of China with grant No. 2016YFA0300401. XR was supported in part by NSFC Nos. 11374024 and 11522429. We thank Yuan Li for providing access to his lab facilities at Peking University. We thank J.~A.~Schneeloch for proof reading.

{\it Note added:} Upon writing up this manuscript, we became aware of an INS work on single crystalline \rucl~\cite{arXiv:1609.00103}.

%\bibliography{qsl}
%merlin.mbs apsrev4-1.bst 2010-07-25 4.21a (PWD, AO, DPC) hacked
%Control: key (0)
%Control: author (8) initials jnrlst
%Control: editor formatted (1) identically to author
%Control: production of article title (-1) disabled
%Control: page (0) single
%Control: year (1) truncated
%Control: production of eprint (0) enabled
%

\end{document}